# Detecting and Directing Single Molecule Binding Events on H-Si(100) with Application to Ultra-dense Data Storage

Roshan Achal[1,2*], Mohammad Rashidi[1,2], Jeremiah Croshaw[1,2], Taleana Huff[1,2], Robert A. Wolkow[1,2,3]

[1]Department of Physics, University of Alberta, Edmonton, Alberta, T6G 2E1, Canada

[2]Quantum Silicon, Inc., Edmonton, Alberta, T6G 2M9, Canada

[3]Nanotechnology Research Centre, National Research Council of Canada, Edmonton, Alberta, T6G 2M9, Canada

[*]Correspondence to: achal@ualberta.ca

**Abstract**

Many new material systems are being explored to enable smaller, more capable and energy efficient devices. These bottom up approaches for atomic and molecular electronics, quantum computation, and data storage all rely on a well-developed understanding of materials at the atomic scale. Here, we report a versatile scanning tunneling microscope (STM) charge characterization technique, which reduces the influence of the typically perturbative STM tip field, to develop this understanding even further. Using this technique, we can now observe single molecule binding events to atomically defined reactive sites (fabricated on a hydrogen-terminated silicon surface) through electronic detection. We then developed a new error correction tool for automated hydrogen lithography, directing molecular hydrogen binding events using these sites to precisely repassivate surface dangling bonds (without the use of a scanned probe). We additionally incorporated this molecular repassivation technique as the primary rewriting mechanism in new ultra-dense atomic data storage designs (0.88 petabits per in$^2$).

**Introduction**

As the end of the current silicon technological roadmap draws closer, ever more capable tools for atomic-scale fabrication are enabling the use of hydrogen-terminated silicon as a platform for a number of alternative avenues[1–7]. The technique known as hydrogen lithography[6,8,9] (HL) has been used on this surface in the creation of atomic-scale logic elements[5], quantum structures[10], ultra-dense rewritable memory arrays[6], and controlled chemical reactions[1,11,12] including the precise placement of dopant atoms[1–3]. Using the tip of a scanning probe microscope, single atoms of hydrogen are removed from the surface to create atomically defined dangling bond (DB) patterns. With HL, tailored reactive sites can even be created for specific molecules[1,11–15] (Supplementary Figure 1). The precise and reproducible spatial integration of molecules into electronic devices is an important consideration towards scalable production, although it remains a considerable challenge. There is a particular focus on the integration of molecules on silicon surfaces to complement and enhance existing technologies[16–20]. This is because molecules can exhibit an array of properties, adding specific functionalities to a given device[16,17,21–24]. The reactivity of a variety of molecules with the silicon surface has been studied, including simple molecules like hydrogen[13,14], and more complex molecules such as alkenes[15,17,20] and phosphine[25,26].

To propel the spatially controlled integration of molecules with the silicon surface forward, new tools that are capable of uncovering atomic-scale details of single molecule reactions are required. Typical scanning tunneling microscope (STM)-based dynamics studies rely on the observations of the motion and state of molecules, or DB sites, to gain insight into their reactivity. This approach is somewhat limited as the smallest intervals between scans are often many seconds to minutes apart[27,28], making real-time observations difficult. There is also the impact of the STM tip as it scans over the area of interest, which can inadvertently deposit material, or strongly influence local dynamics through electric field effects and the injection of charge[28–32]. The latter aspects can additionally complicate the characterization of the amount of charge in DB structures when using only STM measurements, although it is possible to determine the exact charge of defects in some systems through careful analysis and comparison to theory[33,34].

Here, we report an all-STM method, which incorporates HL and hydrogen repassivation (HR) techniques[6,35,36] to readily characterize the total charge of DB structures at the single electron level, with reduced influence from the STM tip. With this STM method we were able to reproduce results[5,37] taken at zero bias with an atomic force microscope (AFM). Once we characterized the number of charges in a given DB structure, such as in an atomically defined reactive site, we then extended this technique to the detection of externally induced charge changes in the DB site and surrounding area. We showed that a single molecule binding event occurring at the DB site can be electronically detected by monitoring for changes in charge. The event can be detected with temporal resolution up to real time, with the possibility of observing multiple bonding events at different DB sites. Combining the ability to precisely create tailored reactive DB sites using HL and the ability to detect a subsequent binding event at those sites opens a new route to studying single molecule reactions, and to test the reliability of theoretically predicted pathways[38–42].

In this work, we also further the prospect of the scalable fabrication of atomic electronics and ultra-dense room-temperature stable memory on hydrogen-terminated silicon. The recent discovery of HR[6,35,36] to complement automated HL[6] has already resulted in significant fabrication advances[5,6,37], although there is still room to improve repassivation speeds. Currently, the only method to controllably add hydrogen to the surface is to sequentially repassivate DBs with atomic hydrogen attached to a scanned probe[6,35,36]. When the probe is depleted of hydrogen, it must travel to gather more. By using HL to create specific DB sites we have demonstrated that we can precisely direct where hydrogen molecules react on the surface of both hydrogen-terminated and deuterium-terminated silicon

(Supplementary Figure 2) to repassivate DBs without a probe, while leaving other DB structures unreacted. This new technique is not only simpler, but also faster than HR because it is unencumbered by the finite number of hydrogen atoms that can be adsorbed to a probe[6,35,36], resulting in a more convenient tool to repair fabrication errors in HL. We then integrated this improved repassivation method as the primary rewriting process in a new proof-of-concept atomic memory array, with a maximum storage density of 1.36 bits per nm$^2$.

**Results and Discussion**

**STM Charge Characterization of Atomically Defined Structures**

Single DBs on an otherwise hydrogen-passivated silicon surface introduce an isolated electronic state within the silicon band gap[10,31,43]. The current through a DB, as measured by an STM tip, can be influenced by the DB's local electrostatic environment[43,44], including by the charge state of subsurface dopants[43]. Due to the sample preparation method there is a dopant-depleted region extending over 60 nm from the surface, which largely isolates surface DBs and dopants in this layer from the bulk[45]. At 4.5 K, dopant atoms laterally separated from the tip by up to 15 nm and at a depth of approximately 5 nm to 15 nm remain un-ionized (neutral) until a critical tip voltage is reached[43]. When one such dopant is field ionized by the tip, the then positive ion core causes downward bending of the local energy bands, thereby creating a conduction channel between the bulk silicon conduction band and the DB level, resulting in a measurable increase of current to the STM tip[43]. This sudden onset of current manifests as a sharp step in the current-voltage, I(V), spectrum taken over the DB[43] (Figure 1). The step is commonly observed in the I(V) spectra of most DBs, however, the exact strength and critical value of this signature depends on the random proximity of the dopant to the surface and a given DB[43]. By controllably adding local negative charges on the surface, so as to introduce upward band bending, it is possible to increase the magnitude of the critical voltage where this onset occurs, therefore requiring a larger tip field to achieve dopant ionization. We take advantage of this effect to develop an STM procedure to characterize the amount of net charge in fabricated DB structures (Figure 1a-e). This was previously achieved with sensitive AFM frequency shift measurements of the charge state transitions of a sensor DB itself[37], or by using an AFM to map the spatial localization of charge in DB structures[5].

It has been predicted that two closely spaced DBs (< 1 nm) on the surface of highly arsenic-doped hydrogen-terminated silicon will share a net charge of one electron, whereas two isolated DBs will have a charge of one electron each[46]. While recent AFM experiments have verified the single net electron occupation of two closely spaced DBs[5,37], the analogous capability has been lacking in STM. This is principally because the STM applies a large perturbative field while imaging and is capable of injecting or removing charge on the surface. By working with sharp tips (approximately 5 nm radius) at voltages between -1.2 V to -1.6 V and characterizing structures in excess of 5 nm laterally removed from the STM tip, the possible effects of both charge injection and field perturbations can be greatly reduced (see Methods).

To determine the net charge of a DB structure with an STM, we first recorded a baseline reference I(V) spectrum over a sensor DB (DB1) (Figure 1a,f-blue), which was selected because it exhibited the sharp current onset due to the ionization of a nearby isolated arsenic dopant. We then added a second isolated DB (DB2) 5.4 nm away (Figure 1b), and a new spectrum was taken over DB1. At this distance, AFM experiments performed with zero applied voltage have shown that DB2 will have a net charge of one electron[5,37]. In this STM measurement, we observed that the presence of DB2 shifted the critical value of the onset of current in the I(V) spectrum of DB1 to a larger negative voltage by -0.06 V (Figure 1f-dark green). The direction of the shift indicates that DB2 is negatively charged. Another DB (DB3) was then added 0.768 nm from DB2 (Figure 1c), and the spectrum of DB1 was measured once more (Figure 1f-light green). The presence of DB3 did not shift the spectrum this time, as would be expected if another negative charge was added into the area. We then erased DB3 using HR and created a new DB 1.15 nm from DB2 (Figure 1d). With this new placement, the spectrum taken over DB1 (Figure 1f-orange) showed an additional shift of -0.065 V, indicating the presence of another charge of one electron in the local area. Repeating this process, DB3 was positioned 1.92 nm from DB1 (Figure 1e), and no additional shift of the I(V) spectrum of DB1 was observed (Figure 1f-red). Since the induced shifts from one and two electrons were very close in magnitude in the I(V) spectrum, we surmise that the un-ionized dopant was sufficiently deep, such that the change in lateral separation of the two DBs in the structure (DB2 and DB3) did not alter their distance to the dopant significantly. Through these STM measurements, we can conclude that 0.768 nm separating DB2 and DB3 gives a net charge of one electron within the pair (due to inter-electron repulsion[46]), while when the spacing is increased to 1.15 nm or more there is a net charge of two electrons within the structure/local area. These results are in agreement with recent AFM studies[5,37] (also see Supplementary Figure 3 for additional AFM results).

Using HL and HR to create and erase a number of isolated DBs (with a charge of one electron) in a particular area, we have demonstrated the capability to calibrate the shifts in the I(V) spectrum of a sensor DB in order to characterize the amount of net charge in larger DB structures, with reduced tip field effects. This technique presents a new opportunity for performing minimally perturbative studies of charge occupation using only an STM (Supplementary Figures 4, 5). In variable-temperature scanned probe systems, we expect it will be possible to observe the I(V) step over a range of temperatures up to approximately 40 K, above which the dopants in the depleted region begin to thermally ionize[47]. Using this technique, we have determined that two immediately adjacent DBs (as shown in Figure 2), forming an inter-dimer site (IDS), have a net charge of one electron among them (Supplementary Figure 4).

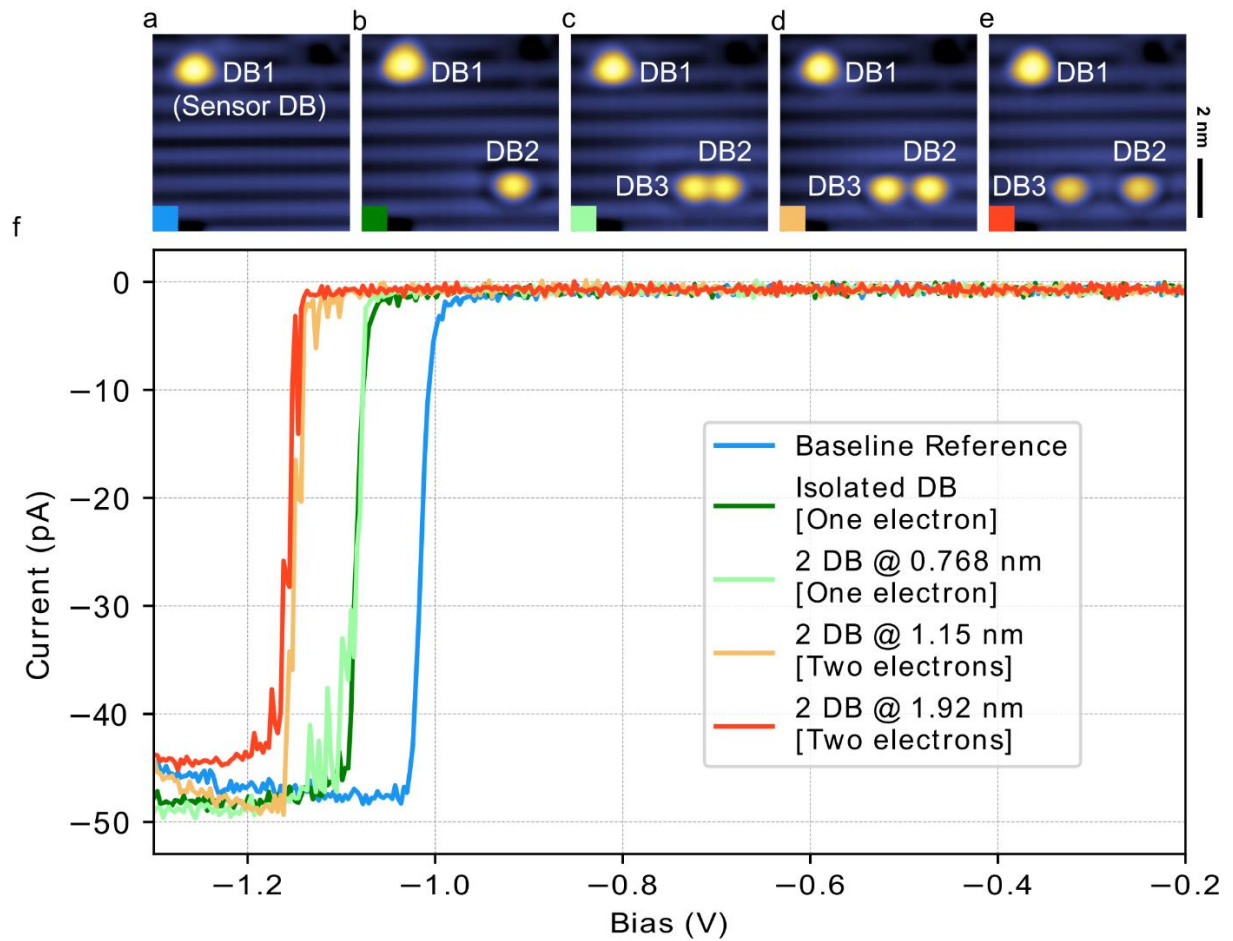

**Figure 1. Characterizing Charge Occupations (V= -1.6 V, I= 50 pA, T= 4.5 K, 6.4 × 6.4 nm².)**

**a)** A scanning tunneling microscope (STM) image of a dangling bond (DB) on the hydrogen-passivated Si(100)-2x1 surface. The DB (DB1) exhibits a sharp current onset in its I(V) spectrum (**f-blue**) due to the ionization of a subsurface arsenic dopant atom caused by the STM tip field. **b)** A second DB (DB2), containing a net charge of one electron is added to the surface 5.4 nm away from DB1, causing the step in the I(V) spectrum of DB1 to shift to the left (**f-dark green**). **c)** A third DB (DB3) is added near DB2, no shift in the I(V) spectrum of DB1 is observed (**f-light green**). **d,e)** The distance between DB2 and DB3 is varied to determine the net charge in the structure for each case. **f)** The I(V) spectra taken over DB1, associated with **a-e**, showing the sharp onset of current.

**Detection of Single-molecule Binding Events**

Once the amount of charge in a particular reactive site (like the IDS) is known, it is possible to use the techniques described above to electronically detect a molecular binding event by periodically recording the I(V) spectrum of a laterally-removed sensor DB (or, to achieve greater time resolution, rapidly sampling the current through it at a fixed bias voltage). If there is a change of at least one electron locally when a molecule binds, then there will be an associated change in the I(V) spectrum (or current). While hydrogen molecules are generally found to be unreactive toward the clean silicon surface[14,20], they are extremely reactive with IDSs on an otherwise hydrogen covered surface[14]. The sites have zero net charge after hydrogen molecules dissociatively adsorb, repassivating their constituent DBs with hydrogen atoms. This is because the two DB states are eliminated upon each DB reforming a bond with a hydrogen atom.

Previous studies were in disagreement over the preferred pathway (inter-dimer[48] vs. intra-dimer[49]) for the dissociative adsorption of hydrogen molecules on the silicon surface. It was eventually shown that the inter-dimer pathway was the dominant one[13], corroborated by theoretical calculations, which predicted it to be barrierless[38–40,42]. These results relied on sequential STM observations of random adsorption events on the surface[13]. With HL, instead of observing random events, we created both the inter- and intra-dimer sites in the same area on the hydrogen-terminated surface. We observed that only the IDSs reacted with hydrogen molecules, providing additional, very direct support for the dominance of the inter-dimer pathway (Supplementary Figure 6).

Figure 2 shows the electronic observation/detection of a single hydrogen molecule binding event to a fabricated IDS. To detect the event, we created an IDS 10.2 nm (thereby reducing any STM field effects) from a sensor DB exhibiting a sharp onset in its I(V) spectrum (Figure 2a,b,e). We then controllably introduced hydrogen gas ($H_2$) into the vacuum chamber to establish a pressure of $4 \cdot 10^{-7}$ Torr and recorded a spectrum over the sensor DB at 35 s intervals (see Methods). After 875 s elapsed, the step in the I(V) spectrum was detected to have shifted back to its original position (Figure 2e), indicating that a binding event occurred, and a hydrogen molecule had dissociatively reacted with the IDS. Subsequent imaging of the reaction site confirmed the DBs of the IDS had been repassivated with hydrogen (Figure 2c). This technique has the potential to be extended to observe multiple reactive sites, with each site shifting the step of a sensor DB an additional amount (Supplementary Figure 7).

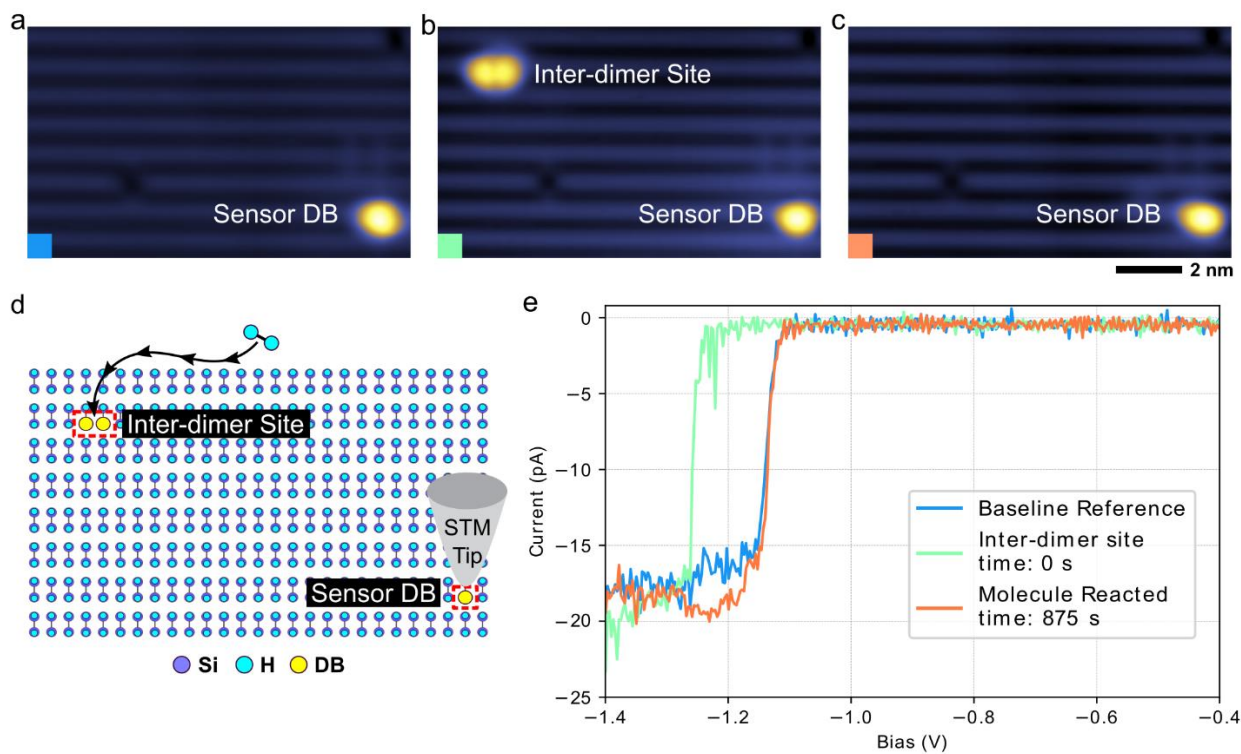

**Figure 2. Electronically detecting a binding event (V= -1.6 V, I= 50 pA, T= 4.5 K, 11.3 × 7 nm².**

**a**) Scanning tunneling microscope (STM) image taken of the area around a sensor dangling bond (DB) exhibiting a sharp current onset in its I(V) spectrum. **b**) STM image after the creation of an inter-dimer site 10.2 nm from the sensor DB. **c**) STM image after hydrogen gas was added into the vacuum chamber, and the reaction of a hydrogen molecule with the inter-dimer site was electronically detected by a shift in the I(V) spectrum taken over the sensor DB. **d**) The geometry of the surface, showing the sensor DB (dashed red square), and the location where an inter-dimer site has been created from DBs (dashed red rectangle) to react with an ambient hydrogen molecule (light blue). **e**) The I(V) spectra taken periodically over the sensor DB associated with **a,b,c**.

**Directing Single-molecule Binding Events (Molecular Hydrogen Repassivation)**

We explored the relative reactivity of hydrogen molecules with a variety of DB structures on both a hydrogen-terminated and deuterium-terminated surface, including with IDSs positioned in situations relevant to HL (with different proximities to other DBs and defects) (Supplementary Figures 2a,b, 7a-e). We observed a lack of reactivity of non-inter-dimer structures with hydrogen molecules (Supplementary Figures 1, 2a,b, 6). We also did not observe any prohibitive effects on the reactivity of the IDSs during our chosen timescales, instead finding them to be highly reactive, even when close to or part of other multi-DB structures. The reactions occurred even while the tip was fully withdrawn from the surface (Supplementary Figure 7). This robust nature and the speed of reaction of hydrogen molecules with IDSs, at an approximate pressure of $1\cdot10^{-9}$ Torr (see Methods for estimation), allowed us to use spatially controlled chemistry to realize a new faster and simplified method for both error correction in automated HL and rewritable binary data storage.

Recent advances in HL have enabled the erasure/repassivation of isolated DBs by bringing in individual hydrogen atoms bonded to a probe tip[6,35,36]. These HR techniques have led to the demonstration of rewritable ultra-dense information storage[6] and atomic circuitry[5]. One limitation of these techniques, however, is that they require the tip to gather hydrogen atoms from locations outside of the fabrication area whenever the tip becomes depleted of available hydrogen. This can slow the fabrication process when many sequential corrections are required and is one of the rate-limiting factors in the rewriting speed of the atomic memory arrays[6]. Instead of bringing in external hydrogen atoms on a probe for HR we can now direct the reaction of ambient molecular hydrogen to erase DBs. This technique is also more generally accessible as it relies on the well-established process of removing hydrogen atoms from the surface, which is less restricted by specific tip geometries or materials, to achieve repassivation[6,50].

To initiate spontaneous molecular hydrogen repassivation (M-HR), we first created an additional DB adjacent to a target DB so as to form a reactive IDS (Figure 3). Then, by working at sufficiently high hydrogen gas pressures (approximately $1\cdot10^{-9}$ Torr), both the targeted DB and the one created adjacent to it were spontaneously repassivated with hydrogen when a hydrogen molecule dissociatively reacted with the site, leaving all other DBs unaltered. The IDS in Figure 3c reacted between the acquisition of Figure 3c and 3d, taking less than 52 s (working at even higher pressures can reduce the time further), while the tip was available to perform other tasks. Working at higher hydrogen pressures has not appeared to impact the long-term stability of the structures fabricated from single DBs (Supplementary

Figures 1, 6). Figure 3f shows the same structure in Figure 3e unchanged six days later in an environment of approximately $1 \cdot 10^{-9}$ Torr.

This "create-to-erase" style of error correction eliminates the need to bring in external hydrogen atoms on a probe for the majority of situations where erroneous DBs need to be corrected during fabrication of atomic circuitry. Desired circuit patterns need not employ any reactive pairings of DBs[5], making all other parts of the circuit immune to the corrective $H_2$ exposure. The reduction in tip movements to gather atomic hydrogen realized by implementing M-HR compared to HR will additionally result in a time savings per correction/rewriting operation. M-HR has the potential to reduce the complexity of the machine learning algorithms[51,52] required for the scalable automation of atomic-scale fabrication and of rewriting atomic memory arrays as well. This is because with M-HR, a single step (the removal of a surface hydrogen atom) can now be used for both fabricating and erasing, as opposed to additionally training neural networks to gather and recognize when a tip is loaded with hydrogen.

With M-HR there is also no physical limitation on the amount of hydrogen available for corrections, as more $H_2$ can always be added into the chamber. In situations where it is not possible to convert an erroneous DB into an IDS, previously established HR techniques can still be used to compliment M-HR, providing a more complete and efficient fabrication toolset. Furthermore, the selective reaction of hydrogen with IDSs has been observed at room temperature[13] and above[48], making it a viable tool for hydrogen lithography in non-cryogenic conditions[8].

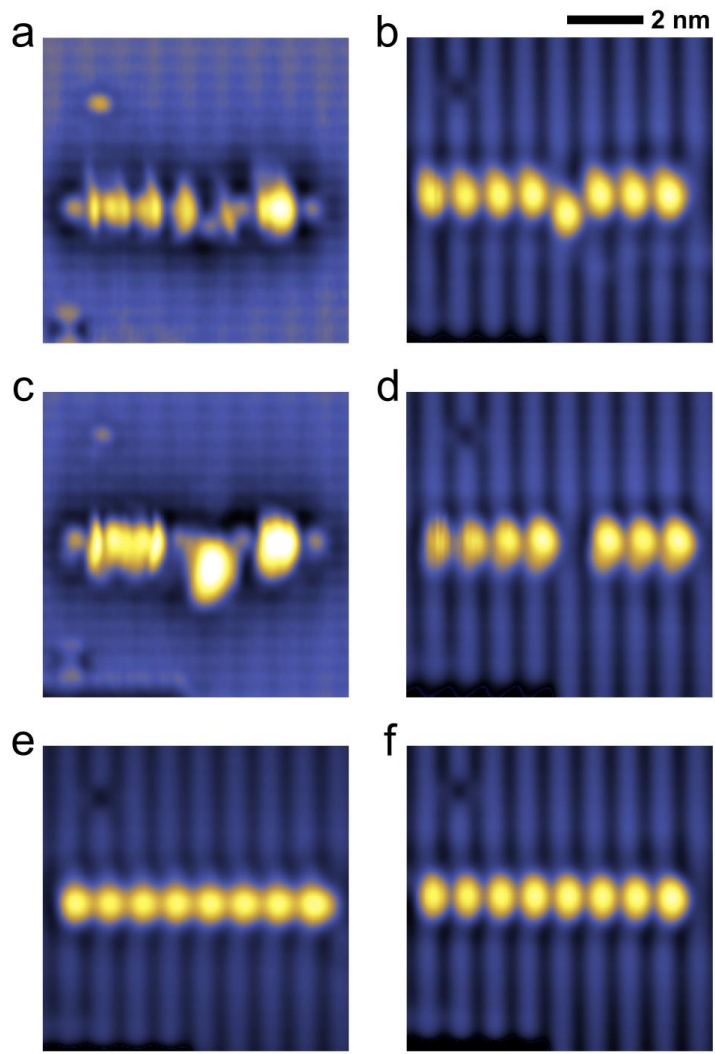

**Figure 3. Molecular hydrogen repassivation (I= 50 pA, T= 4.5 K, 8 × 8 nm$^2$).**

**a)** (V= 1.4 V) A Scanning tunneling microscope (STM) image of an 8-bit dangling bond (DB) memory array, with a fabrication error, taken to visualize the location of the surface hydrogen atoms. **b)** (V= -1.6 V) The same structure in **a**, taken at a negative sample bias to better visualize the location of the DBs in the array. **c)** (V= 1.4 V) An STM image taken after the erroneous DB in **a** was converted into an inter-dimer site, by removing an additional hydrogen atom from the surface. **d)** (V= -1.6 V) The STM image taken immediately after **c** (52 s later), during which time the inter-dimer site reacted with an ambient hydrogen molecule to erase the erroneous DB. **e)** (V= -1.6 V) The error-free 8-bit memory array structure, after the removal of the correct hydrogen atom from the surface. **f)** (V= -1.6 V) The same structure shown in **e** unchanged after 6 days in an environment of an estimated pressure of 1·10$^{-9}$ Torr of hydrogen gas.

**Improved Ultra-dense Atomic Data Storage**

The use of DBs as bits in ultra-dense rewritable atomic memory arrays was recently demonstrated on the hydrogen-terminated silicon surface[6]. Such arrays are a promising candidate for future data storage applications due to the high barriers to diffusion for DBs along the surface, providing stability well above ambient room temperature[53,54]. The primary rewriting mechanism of these arrays is currently HR; however, they can now be redesigned to incorporate M-HR as the main means of altering the stored information (Supplementary Figure 8). In these new designs, each bit/DB can be converted into an IDS to be rewritten as needed, unlike in the original implementation[6] (Supplementary Figure 8a). This alteration in design reduces the maximum storage density from 1.70 bits per $nm^2$ to 1.36 bits per $nm^2$ (Supplementary Figure 8b). The slight reduction in maximum storage density is compensated for by the simplicity and increased maximum speed of rewriting multiple bits, compared to HR (where the tip needs to travel away from the array to be loaded with multiple hydrogen atoms). Additionally, the unlimited local supply of hydrogen molecules removes any restriction on the number of possible write/rewrite cycles of the memory arrays.

We have demonstrated the use of M-HR to rewrite a small 24-bit memory array created using automated HL[6] (Figure 4a,b). With M-HR, once the bits/DBs to be overwritten were converted into IDSs (see Methods), the tip was available to perform other tasks, while the repassivation could proceed in the background in a quasi-parallel fashion. We were able to use the tip to record images in between the acquisition of Figure 4e and Figure 4n, as well as condition it further. This is unlike the HR procedure, which is inherently serial, where the tip is actively involved during the entire repassivation process.

Since the process to replace surface atoms now only requires the technique for atom removal, techniques like M-HR provide a possible path forward from purely scanned probe-based atomic-scale fabrication. In the future, as ion and electron beam-based fabrication techniques become increasingly capable of imaging and manipulating single atoms[55,56], it is conceivable that the STM probe used to remove hydrogen atoms from the silicon surface could eventually be replaced. Should such a transition occur for the removal of atoms, M-HR presents the ability to add material back in a controlled manner, with the prospect of completely scanned probe-free writing/rewriting for data storage applications.

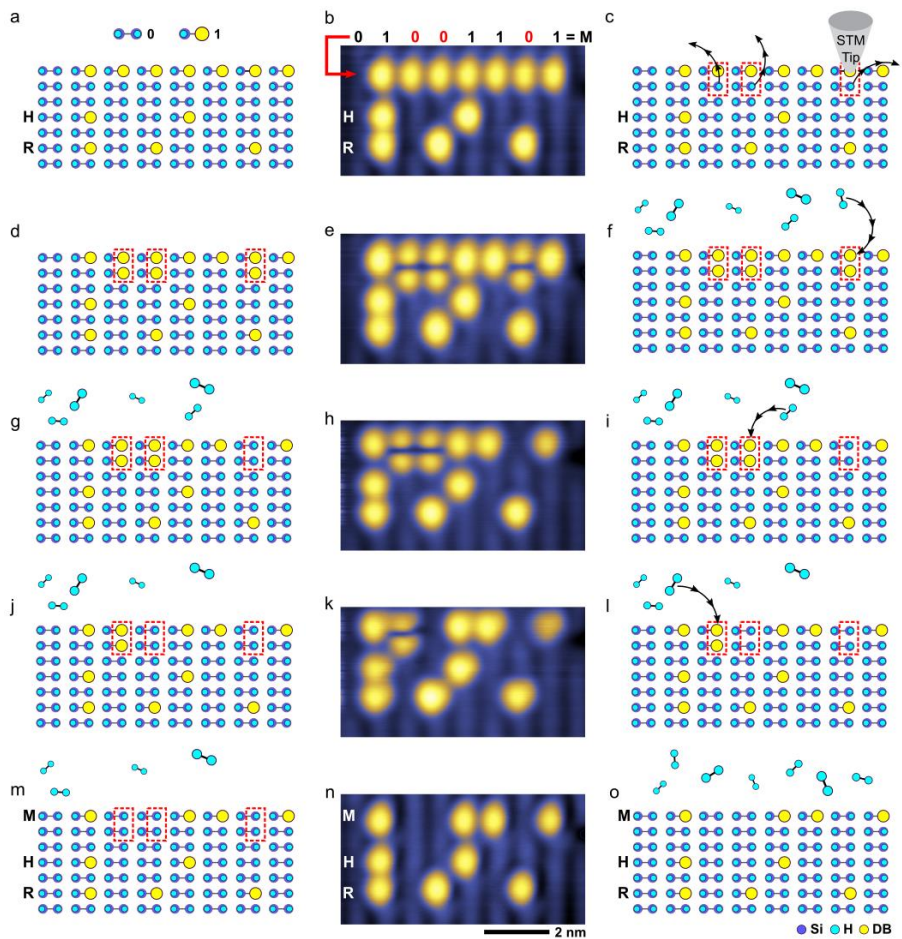

**Figure 4. Rewriting a 24-bit memory array (V= -1.65 V, I= 50 pA, T= 4.5 K, 4 × 7.5 nm$^2$).**

**a**) A schematic of a 24-bit memory array created from dangling bonds (DBs). The first line of the array is blank (01111111), the next two lines are the ASCII binary forms of the letters H (01001000) and R (01010010). **b**) An STM image of the 24-bit array created from DBs (using automated hydrogen lithography[6]) with a storage density of 1.36 bits per nm$^2$. The data in the first line will be rewritten to store the letter M (01001101). **c**) Using the automated scanning tunneling microscope (STM) tip, the surface hydrogen atoms highlighted in red will be removed to create reactive inter-dimer sites in order to rewrite the array. **d**) A schematic of the surface after the three hydrogen atoms shown in **c** have been removed. **e**) An STM image of the 24-bit DB array after three surface hydrogen atoms have been removed with the STM tip to create inter-dimer sites. **f**) Hydrogen gas is introduced into the vacuum chamber to bind with the inter-dimer sites (highlighted in red). **g-i**) Schematics and associated STM images with each molecular hydrogen repassivation event. The first event took 30 minutes to occur while the system reached a working pressure of 1·10$^{-9}$ Torr of hydrogen gas. The following events occurred within a minute of each other. **m**) The first line of the array shown in **a** has been rewritten to the letter M, now that hydrogen molecules have dissociatively reacted with the inter-dimer sites shown in **d**. **n**) An STM image of the 24-bit memory array after the inter-dimer sites have reacted with hydrogen molecules, rewriting the stored information. **o**) The remaining hydrogen gas in the chamber does not react with the isolated DBs in the array and can be used in further rewriting operations.

**Conclusions**

We have demonstrated a process that can be used to conveniently characterize the charge occupation, or changes, of fabricated DB structures down to a single electron level, with only an STM. Using HL and HR techniques to create and erase single isolated DBs with a charge of one electron, the shifts of a feature in the I(V) spectrum of a sensor DB can be calibrated to compare against those induced by a structure of interest. With this technique we verified prior AFM results with an STM.

We electronically detected the binding event of a hydrogen molecule at a prepared IDS on the surface, with the tip laterally removed by > 10 nm. We expect that the same techniques employed here can also be applied to study the adsorption dynamics of additional molecules of technological interest, including diverse alkenes and aromatic molecules by using other tailored DB reactive sites (such as in Supplementary Figure 1). The inter-dimer and intra-dimer sites presented here can directly be used to study cycloaddition reactions on silicon[17], including with ethylene[15,57]. Due to their charged nature, reactive DB sites can also be directly integrated into field-controlled atomic electronic circuitry designs[5], providing yet another route for sensing applications. The ability to exactly position reactive sites in a particular area further opens the possibility of studying the effects of atomic-scale surface variations on the reactivity of otherwise identical DB sites.

We applied the ability to create selective DB sites tailored to react with hydrogen molecules as a new, more efficient, means to correct fabrication errors in automated HL. The unified technique of atom removal and replacement offers several improvements over HR, including no longer requiring external hydrogen atoms to be brought in on a probe. M-HR was then incorporated into new designs of atomic memory arrays to improve the future rewriting speeds and overall usability of atom-based data storage, illustrated with a small-scale demonstration. Although this demonstration only contained 24 bits, there are no physical limitations preventing the technique from scaling to larger arrays.

## Methods

### Equipment

All measurements were performed with a commercial low-temperature Omicron LT-STM (or LT-AFM) operating at 4.5 K. Polycrystalline tungsten wire (0.25 mm diameter) was used for the STM tips. The tips were electrochemically etched in a solution of 2 M NaOH, then were processed under ultra-high vacuum (UHV) conditions in a field ion microscope to further sharpen them via a nitrogen gas etching process[58].

### Sample preparation

The highly arsenic-doped Si(100) (0.003-0.004 ohm-cm) samples were degassed at 600 ˚C under UHV conditions for 24 h. Using resistive heating, the samples were brought to a temperature of 1250 ˚C three to five times via rapid flashes in order to remove all native oxide. We then exposed the samples to $1 \cdot 10^{-6}$ Torr of 99.999% pure hydrogen gas (or 99.7% pure deuterium gas), flowed through a liquid nitrogen trap. A nearby tungsten filament held at 1900 ˚C was used to crack the gas into its atomic constituents. The samples were exposed to the gas for 120 s without heating, then were rapidly flashed to 1250 ˚C. The temperature was then quickly brought down to 330 ˚C for 150 s, giving the hydrogen(or deuterium)-terminated 2x1 surface reconstruction.

### Reducing Tip Field Effects

For sharp tips/probes, with radii of less than 20 nm, there is a significant reduction in the strength of the local tip field and charge injection/extraction along the surface with increasing lateral separation from the tip[10,31,59–63]. Experimentally, on this substrate, the effects of charge injection from an STM tip into a DB have been observed with lateral separations of approximately 2-4 nm depending on the tip radius[10,31,63], with sharper tips requiring closer tip-DB separations for charge injection to occur[31,63]. The effects of the tip field, without the injection of charge, have been observed at up to 7 nm of lateral tip-DB separation, depending on the tip geometry and applied voltage[10,31,37,63]. By working with single atom tips and restricting voltages to between -1.2 V to -1.6 V, and additionally characterizing structures five or more nanometers laterally removed from the STM tip, we greatly reduce the possible effects of both charge injection and field perturbations.

**Electronic Molecular Detection**

Once a sensor DB was identified, a baseline I(V) spectrum was recorded, and the desired number of IDSs were created. The STM tip was then positioned over the sensor DB and the measurement program was initiated to periodically record the I(V) spectrum. We then introduced 99.999% pure hydrogen gas into the system until a pressure of 4·10$^{-7}$ Torr was achieved, via a manual leak valve (the initial base pressure inside of the STM was 5·10$^{-11}$ Torr). The time interval was selected such that the entire I(V) spectrum of the sensor DB could be recorded (both forward and backward sweeps), and the hydrogen gas pressure could be manually corrected in between measurements. The time window also provided sufficient delay for the majority of the mechanical vibrations of the STM tip (due to the manual actuation of the leak valve) to dissipate before each spectrum was acquired. To undertake real-time measurements, where the acquisition of the full I(V) spectrum is not required, the tip can be held at a fixed voltage and the tunneling current (or z-height) can be monitored for sudden changes in the DB conductivity (Supplementary Figure 9). The jump in the tunneling current can be detected in an identical way to the detection of current changes used in HL, with a maximum sampling rate of 10 kHz[6].

**Estimating Pressure**

The time for a hydrogen molecule to bind to an IDS depends on the pressure of hydrogen gas at the sample surface. Due to the design of the STM chamber, the measured pressure outside the cryo-shielding is not an entirely accurate representation of the pressure in the vicinity of the tip and sample. There is low flux of gas through the shielding and the cold shield surfaces lower the pressure further through cryo-pumping effects. Using the observations of the reaction time of these sites, we can estimate a bound on the pressure of hydrogen inside the shielding, where there is no pressure sensor available. To estimate the hydrogen pressure in the vicinity of the tip and sample ($P_{est}$) we assume a perfect sticking coefficient due to the barrierless nature of the reaction of molecular hydrogen with IDSs[38–40,42] (along with our observation of the reaction at 4.5 K). Using the ideal gas law, and the thermal distribution of the velocities and number of hydrogen molecules incident on the silicon surface, the pressure is given by[64]:

$$P_{est} = \frac{d \cdot (2\pi \cdot m_{H_2} \cdot k_b \cdot T)^{\frac{1}{2}}}{t_{rx}}, \qquad (1)$$

where $d$ is the number of surface sites per unit area (for Si(100)-2x1: $d \approx 10^{19}$ sites per m²), $m_{H_2}$ is the mass of a molecule of hydrogen, $k_b$ is the Boltzmann constant, $T$ is the temperature (4.5 K), and $t_{rx}$ is the observed time for all sites to react. Using conventional sequential STM image acquisition for slower dynamics, we observed three IDSs react over 45 h (Supplementary Figure 6), giving an estimated pressure of approximately 1·10$^{-12}$ Torr. With observations taken using the electronic detection technique after the introduction of H₂ into the chamber, including Figure 2, we estimated a local pressure of approximately 1·10$^{-10}$ Torr. In Supplementary Figure 7, all three IDSs reacted within 120 s after creating the third IDS. Including this observation, along with Figure 3, we estimated a pressure at that time of at least 1·10$^{-9}$ Torr near the sample surface, recognizing that the reactions may have occurred in less than 120 s in Supplementary Figure 7.

**Rewriting Atomic Memory Array**

Once the bits/DBs identified to be rewritten were converted into IDSs (Figure 4), hydrogen gas was introduced into the vacuum chamber to achieve the same conditions as in Figure 3. Because the initial hydrogen background pressure in the chamber was low (< 1·10$^{-11}$ Torr), it took approximately 30 minutes before the pressure in proximity to the tip and sample was sufficiently high for the first M-HR event to occur. The remaining events occurred within one minute of each other after an approximate pressure of 1·10$^{-9}$ Torr was achieved.

**Acknowledgements**

The authors thank M. Salomons and M. Cloutier for their technical assistance, J. Pitters for helpful discussions, and J. Phillips for proofreading the manuscript. We would like to also thank NSERC, AITF, NRC, and QSi for their financial support.

**Author Contributions**

R.A. conceived of the memory designs, molecular hydrogen repassivation procedure, molecule detection procedure, and performed all STM experiments. M.R., R.A., and R.A.W., conceived of the charge characterization experiment. J.C. and R.A. developed the molecule detection program. T.H. performed the supplementary AFM measurements. R.A.W. supervised the project. R.A. prepared the manuscript. All authors participated in the review and discussion of the manuscript and results.

**Competing Interest**

A patent application is in process related to molecular hydrogen repassivation and memory designs as described in the manuscript. All of the authors are affiliated with Quantum Silicon Inc (QSi). QSi is seeking to commercialize atomic silicon quantum dot-based technologies.

# Supplementary Information

**Detecting and Directing Single Molecule Binding Events on H-Si(100) with Application to Ultra-dense Data Storage**

Roshan Achal[1,2*], Mohammad Rashidi[1,2], Jeremiah Croshaw[1,2], Taleana Huff[1,2], Robert A. Wolkow[1,2,3]

[1]Department of Physics, University of Alberta, Edmonton, Alberta, T6G 2E1, Canada

[2]Quantum Silicon, Inc., Edmonton, Alberta, T6G 2M9, Canada

[3]Nanotechnology Research Centre, National Research Council of Canada, Edmonton, Alberta, T6G 2M9, Canada

*Correspondence to: [achal@ualberta.ca](achal@ualberta.ca)

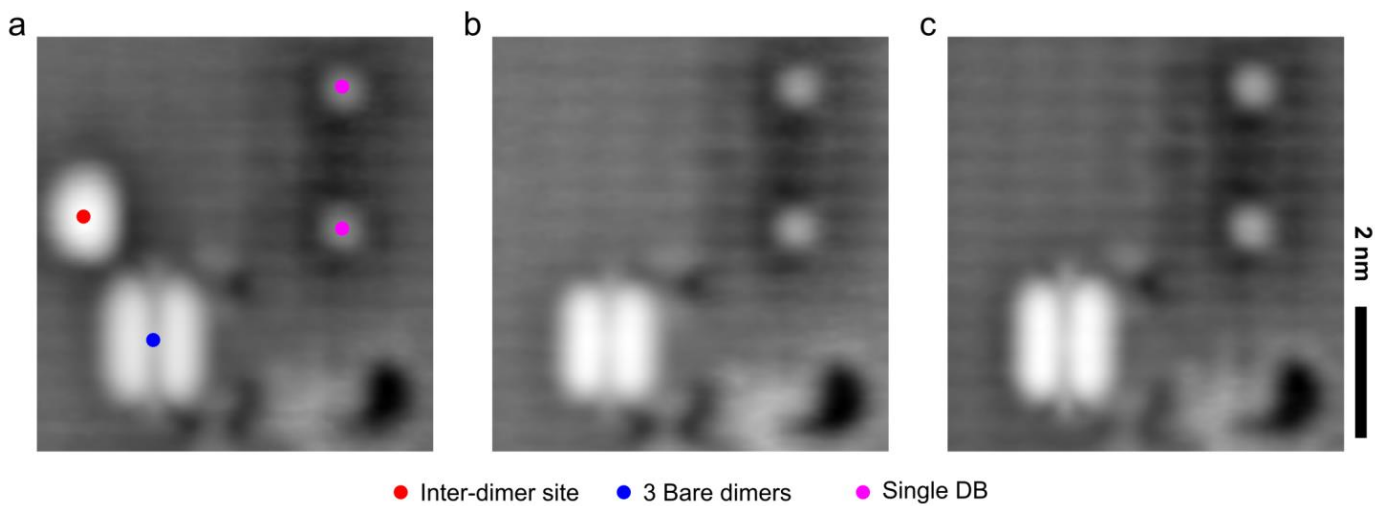

**Supplementary Figure 1. Reactive Sites (V= 1.4 V, I= 50 pA, T= 4.5 K, 6.34 × 6 nm$^2$).**

**a**) A scanning tunneling microscope (STM) image of an area where an inter-dimer site (red) has been created along with a site containing three adjacent bare dimers (blue), and two single dangling bonds (DBs) (pink). The inter-dimer site is highly reactive with hydrogen molecules, while the site with three bare dimers is reactive with phosphine molecules[1]. **b**) An STM image of the same area in **a** after sitting in a vacuum chamber with a base pressure of 5·10$^{-11}$ Torr for 44 hours. In this time, only the inter-dimer site has reacted with a hydrogen molecule, despite the 3 bare dimer site containing areas with equivalent geometry to an inter-dimer site. **c**) An STM image of the same area in **b** after sitting for an additional 97 hours in the vacuum chamber. The cryogens in the STM had to be replenished twice during this time, so the sample was subjected to temperature spikes up to 15 K and pressure spikes of up to 1·10$^{-8}$ Torr during this process. Again, no reactions were observed in this time.

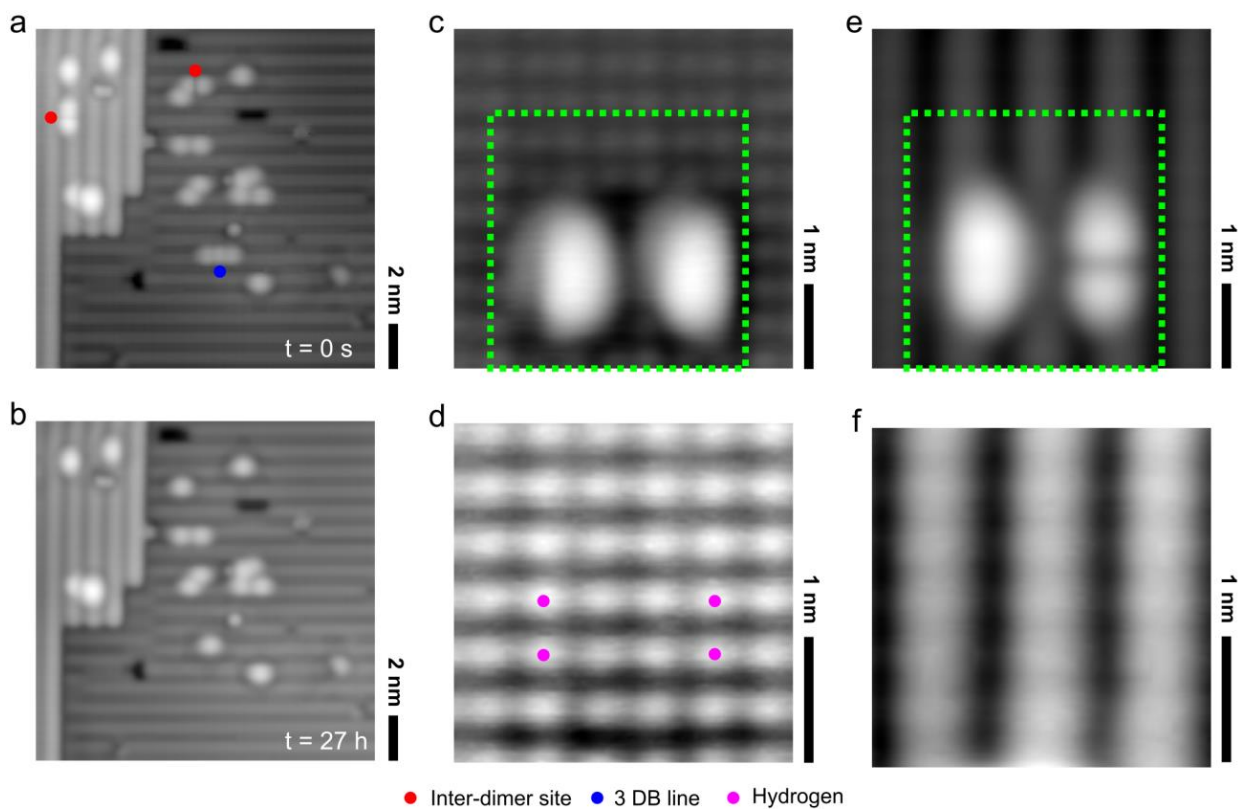

● Inter-dimer site ● 3 DB line ● Hydrogen

**Supplementary Figure 2. Reactions with deuterium-terminated Si(100)-2x1 (I= 50 pA, T= 4.5 K).**

**a)** (V= -1.8 V, 15 × 15 nm$^2$) A scanning tunneling microscope (STM) image of a deuterium-terminated Si(100)-2x1 surface, with various dangling bond (DB) structures. **b)** (V= -1.8 V, 15 × 15 nm$^2$) The same area as in the previous figure after 27 hours in an environment of 9·10$^{-11}$ Torr of hydrogen gas (negligible deuterium background). At each inter-dimer site, a hydrogen molecule has reacted. In the structure with three directly adjacent DBs (containing an inter-dimer site), a hydrogen molecule reacted with two of the DBs, leaving an isolated DB remaining. **c)** (V= 1.4 V, 4 × 4 nm$^2$) An STM image of two inter-dimer sites created on the deuterated surface, ready to react with an ambient hydrogen molecule. **d)** (V= 1.4 V, 2.7 × 2.7 nm$^2$) An STM image of the area highlighted in **c** after the reaction of two hydrogen molecules. With the present imaging techniques, the hydrogen and deuterium atoms could not be differentiated. Inelastic tunneling spectroscopy[2] was also unable to resolve a discernible signal to identify the atoms. **e)** (V= -1.8 V, 4 × 4 nm$^2$) Filled states image of **c**. **f)** (V= -1.8 V, 2.7 × 2.7 nm$^2$) Filled states image of **d**.

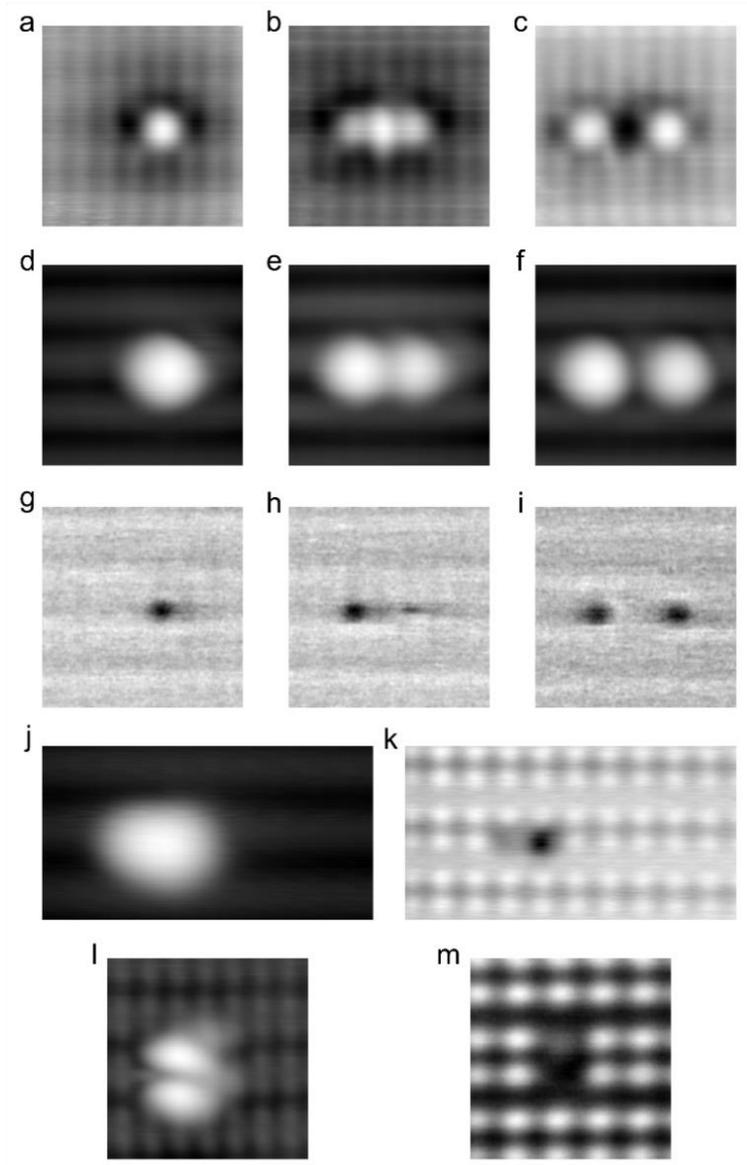

**Supplementary Figure 3. Atomic force microscope measurements of net charge in DB structures.**

**a-c**) (V= 1.3 V, I= 50 pA, T= 4.5 K, 3 × 3 nm$^2$) Scanning tunneling microscope (STM) images of two dangling bonds (DBs) with various separations. **d-f**) (V= -1.8 V, I= 50 pA, T= 4.5 K, 3 × 3 nm$^2$) STM images of the same DBs in **a-c**. **g-h**) (V= 0 V, Z$_{rel}$= -300 pm, T= 4.5 K, 3 × 3 nm$^2$) Constant height atomic force microscope (AFM) frequency shift images of the structures in **a-c**. The dark depressions represent the location of an electron within each structure[3]. In **g** and **h** there is only a net charge of one electron within the structures. In **i** there are two net electrons present. **j**) (V= -1.8 V, I= 50 pA, T= 4.5 K, 2 × 4 nm$^2$) An STM image of an inter-dimer site. **k**) (V= 0 V, Z$_{rel}$= -300 pm, T= 4.5 K, 2 × 4 nm$^2$) Constant height AFM frequency shift image of the inter-dimer site in **j**, showing the presence of only one net electron. **l**) (V= 1.3 V, I= 50 pA, T= 4.5 K, 3 × 3 nm$^2$) An STM image of an intra-dimer site. **m**) (V= 0 V, Z$_{rel}$= -300 pm, T= 4.5 K, 2 × 4 nm$^2$) Constant height AFM frequency shift image of the intra-dimer site (bare dimer) in **l**, showing the presence of only one net electron.

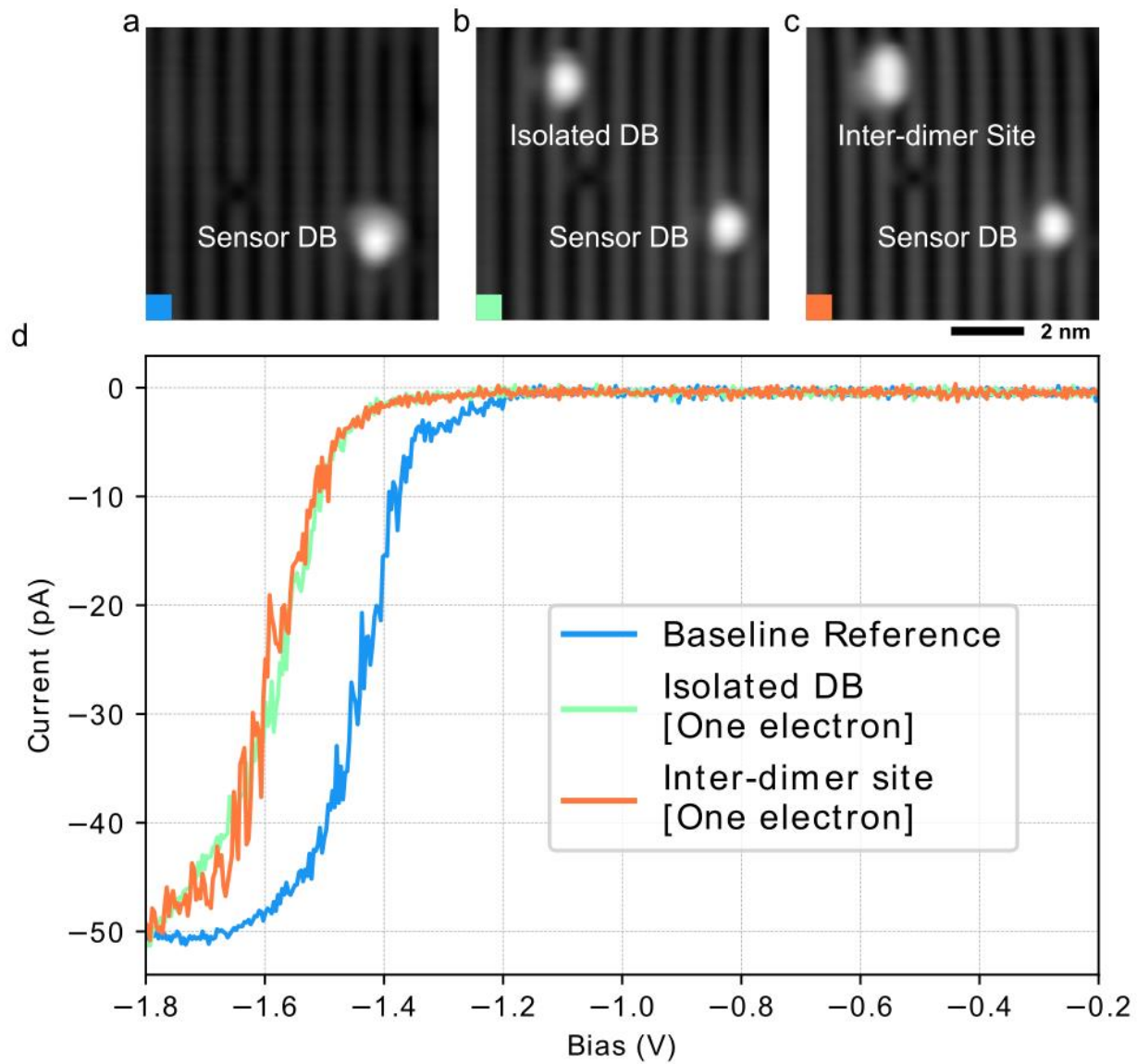

**Supplementary Figure 4. Net charge in an inter-dimer site (V= -1.8 V, I= 50 pA, T= 4.5 K, 8 × 8 nm$^2$).**

**a**) A scanning tunneling microscope (STM) image of a dangling bond (DB) on the hydrogen-passivated Si(100)-2x1 surface. This DB exhibits a sharp current onset in its I(V) spectrum due to the ionization of a subsurface arsenic dopant atom caused by the STM tip field, making it suitable to act as a charge sensor. **b**) A second DB, containing a net charge of one electron is added to the surface to calibrate the sensor. **c**) The DB in **b** is converted into an inter-dimer site. **d**) The I(V) spectra taken over the first DB, associated with **a-c**, showing the shift in the sharp onset of current.

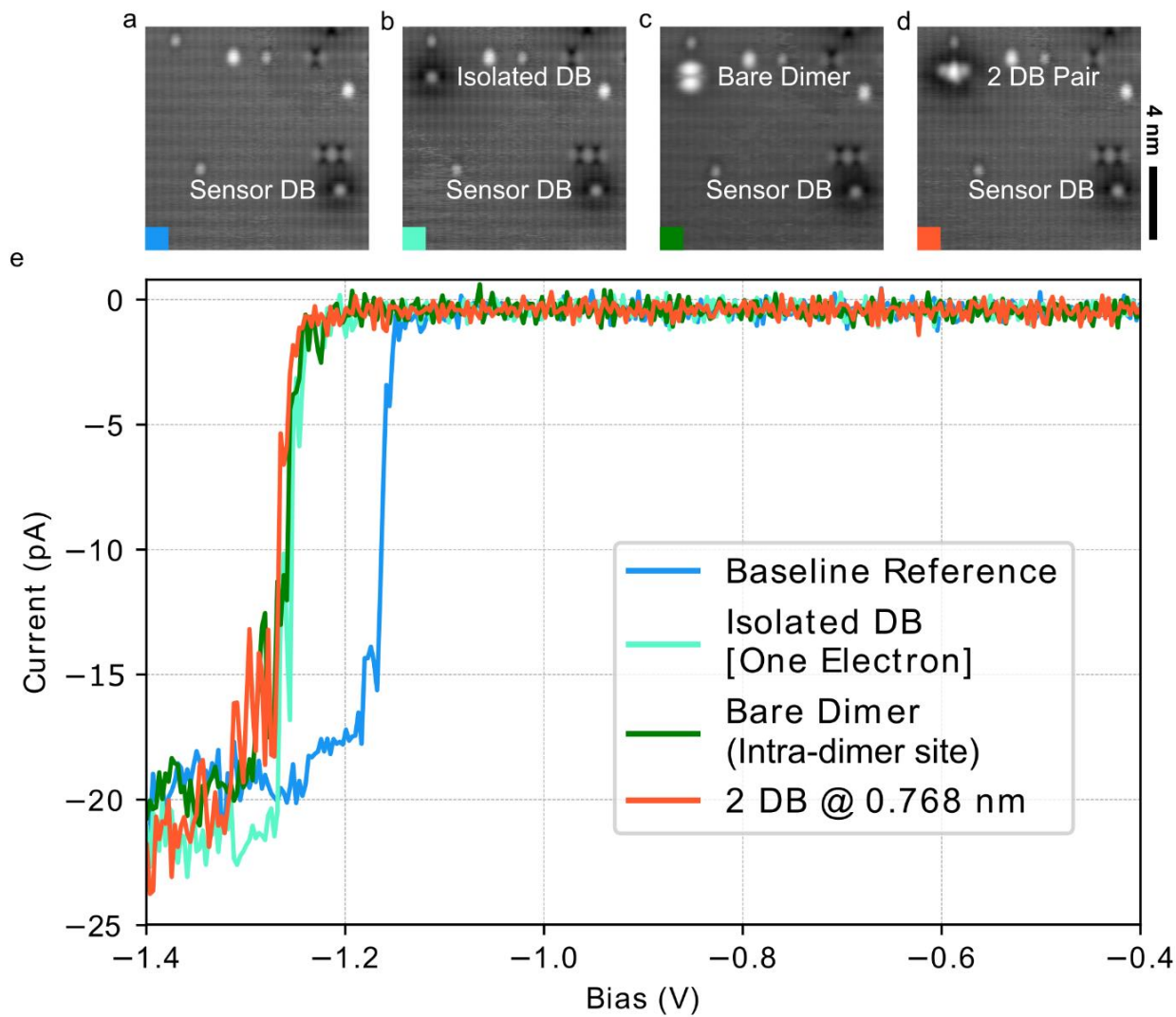

**Supplementary Figure 5. Net charge in other structures (V= 1.4 V, I= 50 pA, T= 4.5 K, 12 × 12 nm$^2$).**

**a-d**) Scanning tunneling microscope (STM) images of the dangling bond (DB) structures associated with the spectra in **e**. **e**) The I(V) spectra taken over the sensor DB in **a-d**. After the shift in the I(V) spectrum of the sensor DB was calibrated in **b**, the structures in **c** and **d** were determined to have a net charge of one electron. These results correspond with the atomic force microscope measurements shown in Supplementary Figure 3.

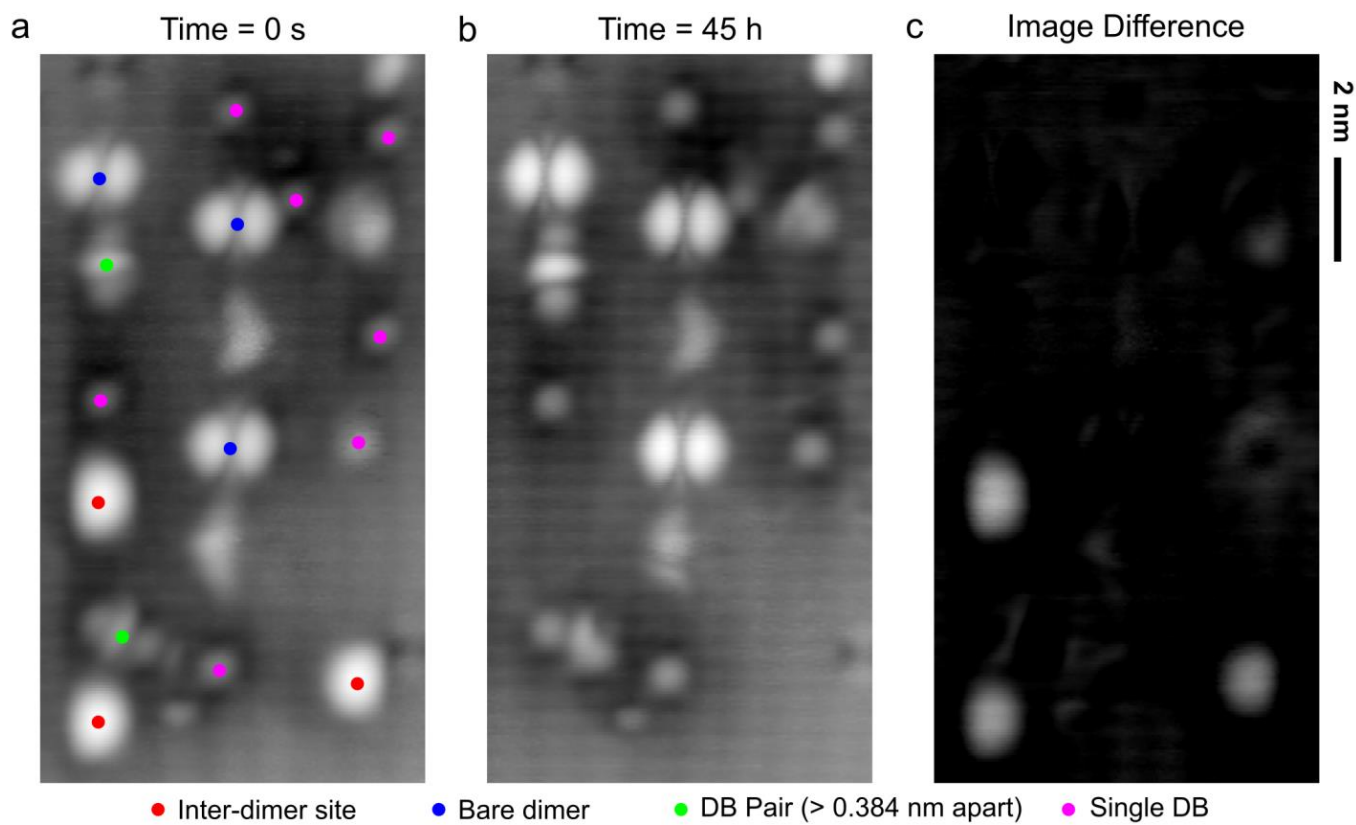

**Supplementary Figure 6. Various dangling bond structures (V= 1.4 V, I= 50 pA, T= 4.5 K, 14 × 8 nm$^2$).**

**a**) A scanning tunneling microscope (STM) image of an area where a number of various DB structures have been created. Inter-dimer sites are denoted in red, while the intra-dimer sites (bare dimers) are denoted in blue. **b**) An STM image of the same area after sitting in a vacuum chamber with a base pressure of 5·10$^{-11}$ Torr for 45 hours. **c**) A difference image between **a** and **b**. Only the inter-dimer sites have reacted after 45 hours, with no other significant changes occurring in this time with any of the isolated DBs or other DB structures.

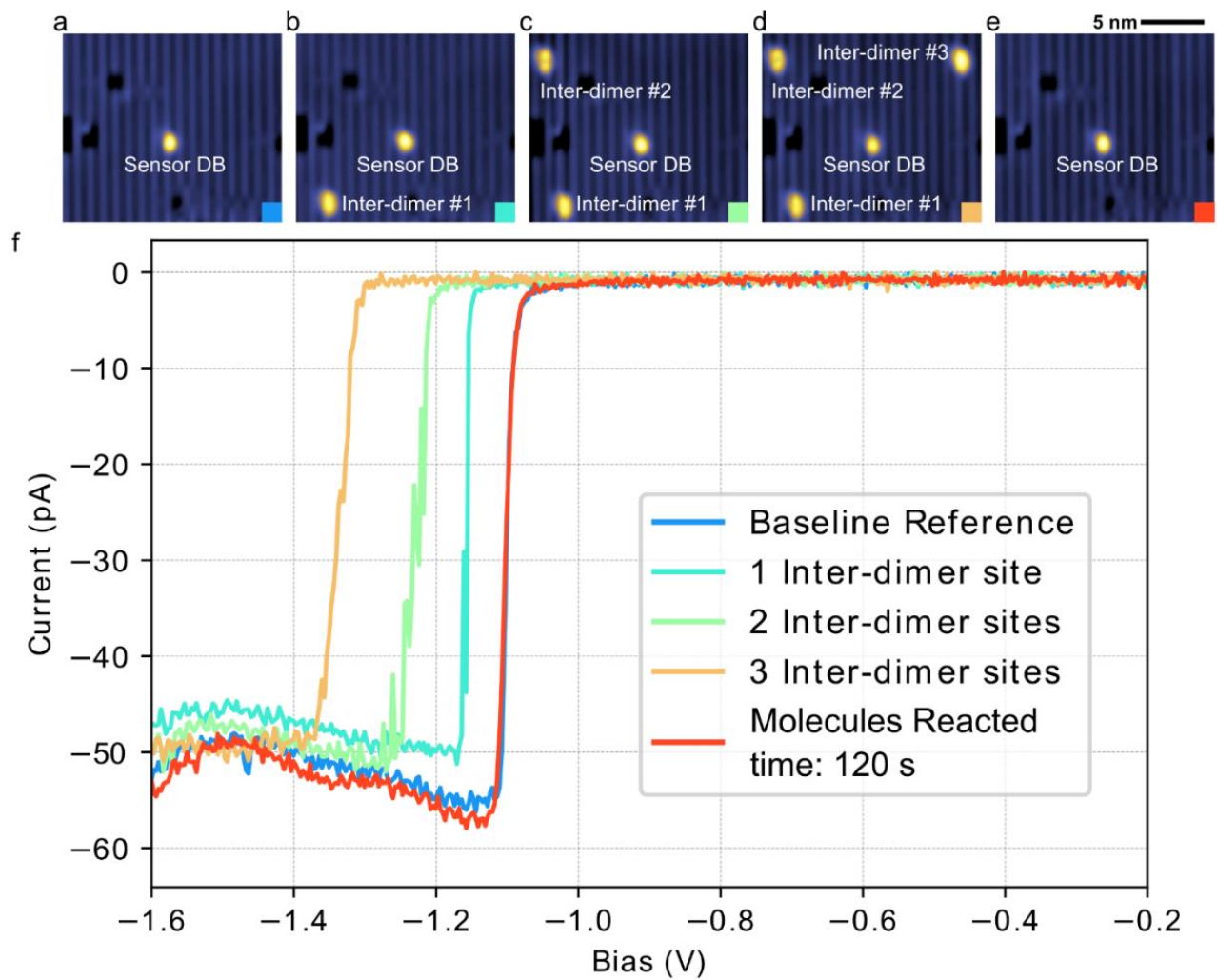

**Supplementary Figure 7. Multiple inter-dimer sites (V= -1.7 V, I= 50 pA, T= 4.5 K, 15 × 13 nm².)**

We created three inter-dimer sites 6.6 nm to 8.9 nm from a sensor DB, with the intention of detecting three sequential binding events. The observation of multiple sites is possible because the presence of each subsequent inter-dimer site shifted the I(V) step used for electronic detection to increasingly negative voltages, as seen in **f**. While the tip was withdrawn from the surface (in preparation of adding additional $H_2$ into the chamber before initiating monitoring), however, all three sites reacted within 120 s. The sites reacted so quickly because the base-pressure near the sample surface remained elevated after earlier trials (estimated to be at least $1 \cdot 10^{-9}$ Torr). With an improved experimental procedure and lower hydrogen background pressure such multi-reaction sequences can be monitored.
**a)** A scanning tunneling microscope (STM) image of an isolated dangling bond (DB) on the hydrogen-passivated Si(100)-2x1 surface, which exhibits a sharp current onset in its I(V) spectrum (**f-blue**). **b-d)** STM images of a sequence of inter-dimer sites being added at various distances and locations relative to the sensor DB (6.6 nm, 8.9 nm, 8.4 nm). **e)** An STM image taken 120 s after the creation of the inter-dimer site in **d**, before any additional hydrogen gas was added into the vacuum chamber. **f)** The I(V) spectra taken over the sensor DB, associated with **a-e**.

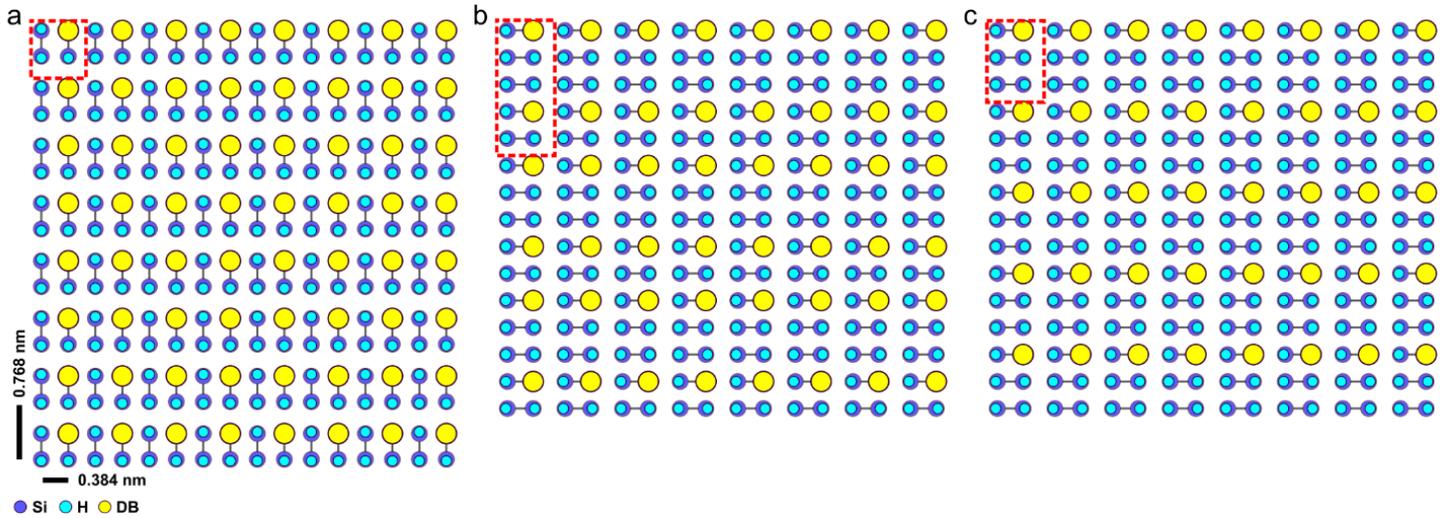

**Supplementary Figure 8. Ultra-dense atomic memory designs (8 bits per line).**

**a**) The original design of the ultra-dense memory array using dangling bonds (DBs) to represent one bit of information[4], with a maximum bit density of 1.7 bits per nm$^2$. The unit memory cell is denoted in red, containing one bit. This memory was designed to be rewritten by bringing in external hydrogen atoms on a probe tip, which presented a potential bottleneck for the speed of rewriting operations. **b**) A new array design for DB-based atomic storage, with a maximum storage density of 1.36 bits per nm$^2$. In this scheme the memory cell denoted in red contains two bits. Each bit within the cell can now be rewritten by converting the DB into an inter-dimer site, which subsequently reacts with an ambient molecule of hydrogen. This removes the need to bring in hydrogen atoms externally. **c**) A less dense array design with a maximum bit density of 1.13 bits per nm$^2$. In this scheme the memory cell denoted in red contains one bit. Here, each bit can also be rewritten through molecular hydrogen repassivation, with a more analogous geometry to **a**.

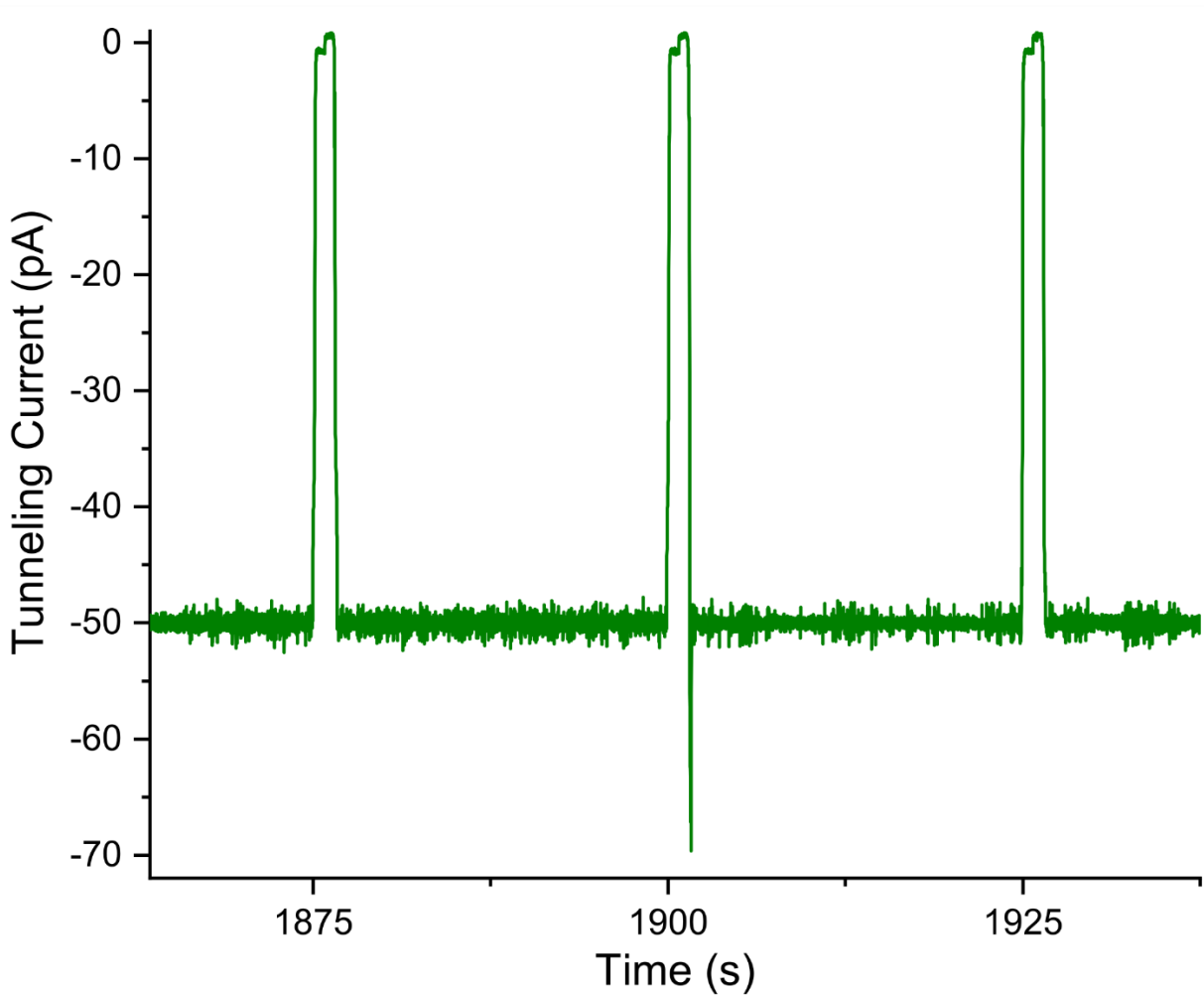

**Supplementary Figure 9. Tunneling current spike during molecular monitoring.**

Live sampling of the tunneling current while periodically monitoring a sensor dangling bond (DB). The periodic I(V) measurements were taken every 25 s. When a hydrogen molecule binds to the monitored inter-dimer site, the conductivity of the sensor DB changes, causing a spike in the tunneling current, before the feed-back control adjusts the tip height to return to the set current. The spike in current just after 1900 s indicates that the monitored inter-dimer site reacted with a hydrogen molecule. For fast time resolution of binding events, the tunneling current can be monitored for spikes, rather than recording I(V) traces periodically.